\documentclass[letterpaper,12pt]{article}
\usepackage{graphicx} 
\usepackage{amssymb} 
\usepackage{empheq}
\usepackage{url}
\usepackage{natbib}
\usepackage{booktabs} % for "toprule", "middlerule" in tables
\usepackage{adjustbox} % sizes flowchart, table,...
\DeclarePairedDelimiter\floor{\lfloor}{\rfloor}

\newcommand{\med}{\mbox{median}}
\newcommand{\RD}{\mbox{RD}}

\newcommand{\LB}{\mbox{LB}}
\newcommand{\diag}{\textrm{diag}}

\newcommand{\ba}{\boldsymbol a}
\newcommand{\bb}{\boldsymbol b}

\newcommand{\bx}{\boldsymbol x}
\newcommand{\by}{\boldsymbol y}
\newcommand{\bz}{\boldsymbol z}
\newcommand{\bdelta}{\boldsymbol \delta}
\newcommand{\bmu}{\boldsymbol \mu}

\newcommand{\bA}{\boldsymbol A}
\newcommand{\bB}{\boldsymbol B}

\newcommand{\bI}{\boldsymbol I}
\newcommand{\bS}{\boldsymbol S}

\newcommand{\bX}{\boldsymbol X}

\newcommand{\bZ}{\boldsymbol Z}

\newcommand{\bSigma}{\boldsymbol \Sigma}

\newcommand{\hbmu}{\hat{\boldsymbol \mu}}
\newcommand{\hbSigma}{\hat{\boldsymbol \Sigma}}

\begin{document}
	
\title{Real-time discriminant analysis in the presence of label and measurement noise} 

\author{Iwein Vranckx, Jakob Raymaekers, 
        Bart De Ketelaere,\\ 
				Peter J. Rousseeuw, Mia Hubert\\ \\
        KU Leuven, BE-3001 Heverlee, Belgium}

\date{November 10, 2020}

\maketitle

\begin{abstract}
Quadratic discriminant analysis (QDA) is a widely used classification technique. Based on a training dataset, each class in the data is characterized by an estimate of its center and shape, which can then be used to assign unseen observations to one of the classes. The traditional QDA rule relies on the empirical mean and covariance matrix. Unfortunately, these estimators are sensitive to label and measurement noise which often impairs the model's predictive ability. Robust estimators of location and scatter are resistant to this type of contamination. However, they have a prohibitive computational cost for large scale industrial experiments. We present a novel QDA method based on a recent real-time robust algorithm. We additionally integrate an anomaly detection step to classify the most atypical observations into a separate class of outliers. Finally, we introduce the label bias plot, a graphical display to identify label and measurement noise in the training data. The performance of the proposed approach is illustrated in a simulation study with huge datasets, and on real datasets about diabetes and fruit.\\
\end{abstract}
		
		
\section{Introduction} \label{sec:intro}	

Supervised classification is a very common task in statistics and machine learning.
Given a training dataset of labeled instances, the goal is to train a classifier such that new observations can be classified into one of the known classes (groups).
Many classification techniques exist, see e.g.~\cite{hastie2009elements,Murphy2012ML,james2013introduction} for an overview. We will focus on discriminant analysis (DA), one of the oldest and well-studied techniques. DA is based on the underlying assumption that the data follow a mixture of multivariate normal distributions. In its basic form DA has several attractive properties, most notably its conceptual and computational simplicity.

Traditional DA relies on the empirical mean and covariance matrix of each class. Despite its nice properties, it is highly sensitive to violations of the mixture density assumption.
Two sources of these violations which commonly occur in practice are \textit{label noise} and \textit{measurement noise}. Label noise, also called mislabeling, occurs when instances in the training data have been given a wrong label, so their recorded label differs from their actual one. See \cite{frenay2013classification} for a comprehensive survey. These instances can heavily affect the classification result, since they essentially encourage the classifier to associate characteristics of one class with the label of another class. As training data is often labeled manually, the labels are prone to human error and some degree of mislabeling is likely to occur in practice.

The second type of noise, measurement noise, occurs when observations in the training data have deviating measurements. Such outliers may affect the mean and covariance matrix of their class which is then characterized poorly, causing classical DA to underperform. 

Several proposals have been made to make DA robust against label and measurement noise, see for example \cite{chork1992integrating,croux2001robust,hubert2004fast}. These methods all rely on robust estimators of location and scatter, which work well but need substantial computation time for large datasets. The context of this paper is that of real-time classification in an industrial setting, such as food sorting or classification of plastics and glass. Typically, huge amounts of product are scanned in an automated inspection process. Robust discriminant analysis is an absolute must in this setting, since these datasets are typically corrupted by both label and measurement noise. Unfortunately, none of the previously mentioned robust algorithms can handle the sheer volume of data that is generated by these classification tasks.

In this paper we address this issue by incorporating the recently introduced RT-DetMCD method~\citep{de2020real}, a real-time robust estimator of location and scatter, into the discriminant analysis framework. We further integrate an anomaly detection step: if an observation does not match any of the known classes it will be classified into a separate outlier class, explicitly revealing significant measurement errors in the training set. The resulting approach allows us to combine a high degree of robustness against label and measurement noise with a low computation time. We also introduce a graphical display to identify label and measurement noise in the training data. Using extensive simulations we show that the proposed approach works well at huge datasets, even with high noise rates. The accuracy of the method is also illustrated on two real datasets, in which we identify and interpret several sets of atypical observations.

The remainder of the paper is organized as follows. In Section~\ref{sec:robda} we describe our real-time robust classifier, which incorporates the anomaly detection step. Section~\ref{sec:LBplot} introduces the new graphical display. The simulation study in Section~\ref{sec:simulation} compares the performance of the proposed method with that of classical discriminant analysis under label and measurement noise. Section~\ref{sec:experiments} illustrates the method on two real datasets. Finally, the main conclusions are summarized in Section~\ref{sec:conclusions}.

\section{Real-time robust QDA}\label{sec:robda}

\subsection{Discriminant analysis}
Suppose we have a $p$-variate random vector $X$ which describes the data generated by an experiment. Assume $X$ follows a multivariate normal mixture model with $G$ classes (subpopulations), i.e.\ the density of $X$ can be written as $f(\bx) = \sum_{g = 1}^{G}{p_g f_g(\bx)}$ where $p_g$ denotes the prior probability of class $g = 1,\ldots,G$ and $f_g \sim N(\bmu_g, \bSigma_g)$ is the $p$-variate normal density of class $g$ described by the location vector $\bmu_g$ and scatter matrix $\bSigma_g$. The aim is to divide the $p$-dimensional space into $G$ regions which correspond to the classes. Based on these regions, new cases can be classified into one of the classes. To find these regions DA uses the Bayes discriminant rule, which assigns an unknown observation $\bx$ to the class $g$ for which $\ln(p_g f_g(\bx))$ is highest among all $G$ classes. Using the density of the multivariate normal distribution one obtains the quadratic discriminant analysis (QDA) rule: assign $\bx$ to the class~$g$ for which the quadratic discriminant score $d(\bx, \bmu_g, \bSigma_g,p_g)$ is highest, with
\begin{equation}
\label{eq:QDA}
d(\bx, \bmu_g, \bSigma_g,p_g) = -\frac{1}{2} \ln|\bSigma_g| - \frac{1}{2} (\bx- \bmu_g)^\top \bSigma_g^{-1} (\bx- \bmu_g) + \ln(p_g).
\end{equation}

If the covariance matrices $\bSigma_g$ of all classes are equal they can be replaced by a common covariance matrix $\bSigma$, which leads to linear discriminant scores and the corresponding linear discriminant analysis (LDA) method. However, as this homoskedasticity assumption is often not realistic in practical settings, we will concentrate on QDA.

\subsection{Classical discriminant analysis}\label{subsec:CQDA}
The quadratic discriminant scores~\eqref{eq:QDA} are computed based on the prior probabilities $p_g$\,, the means $\bmu_g$ and the covariance matrices $\bSigma_g$\,, which all have to be estimated from the data. Suppose we have a multivariate dataset $\bX$ of $n$ observations in $p$ dimensions sampled from $G$ different classes, and a class label vector $\by$ of length $n$ with $y_i$ in $\{1,\ldots,G\}$ for all $i=1,\ldots,n$. Denote by $n_g$ the number of observations from class $g$ in the data. The set of observations and labels $(\bX, \by)$ is called the training set. In order to estimate the unknown parameters $\bmu_g$ and $\bSigma_g$\,, classical QDA (CQDA) uses the empirical mean $\bar{\bx}_g$ and empirical covariance matrix $\bS_g$ of each class. The unknown membership probabilities are estimated using the relative frequencies of each class in the training data: $\hat{p}_{g,C} = n_g/n$ where $C$ stands for classical. The CQDA rule then assigns $\bx$ to the class $g$ for which $d(\bx,\bar{\bx}_g,\bS_g,\hat{p}_{g,C})$ is highest. Note that the data dimension $p$ should be below $\min_g n_g$ since otherwise $\bS_g$ becomes singular.

\begin{figure}[!ht]
	\centering
	\includegraphics[width=\linewidth]{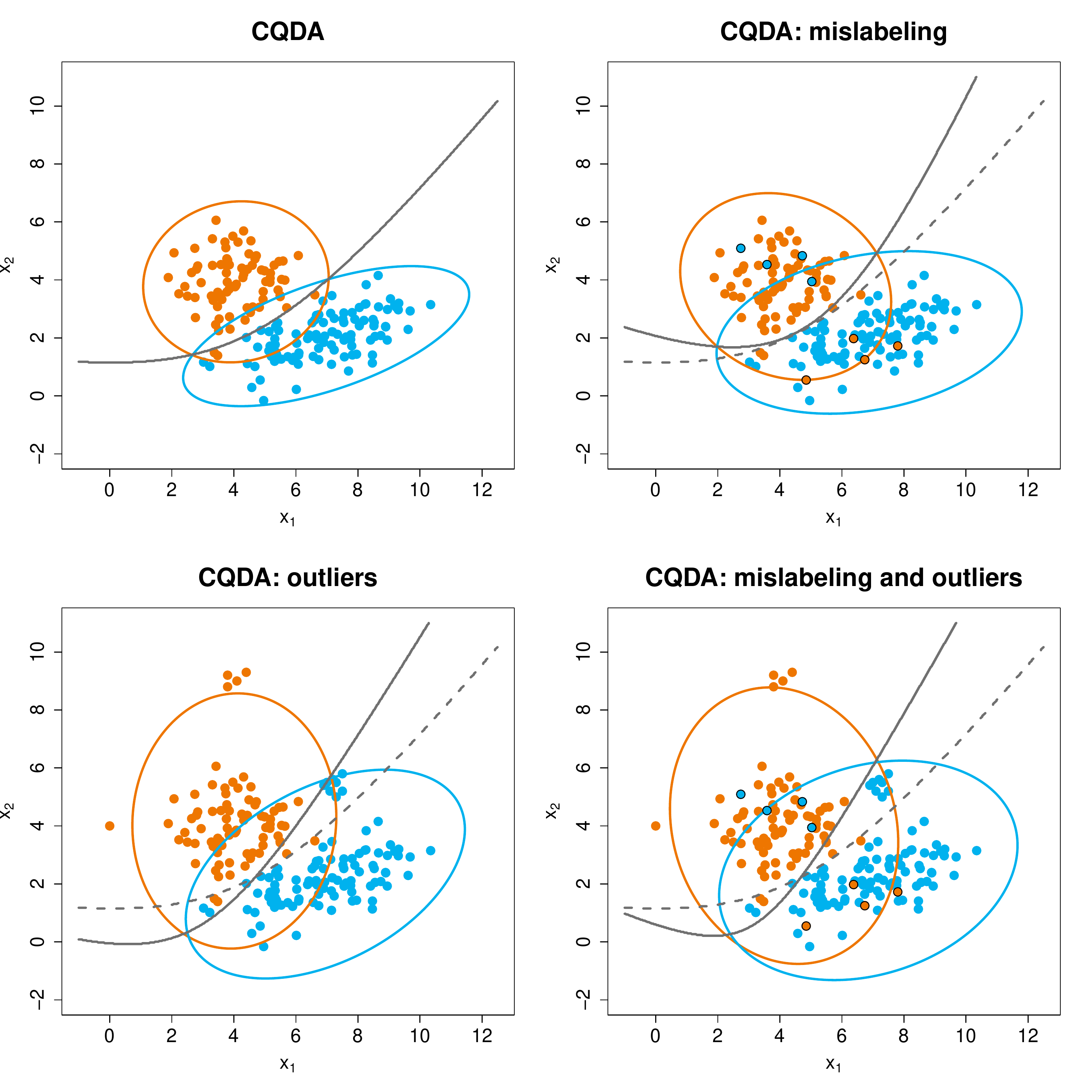}
	\caption{The effect of label and measurement noise on classical quadratic discriminant analysis.} 
	\label{fig:noisyQDA}
\end{figure}

Although CQDA remains a popular classification method, it is known that it is very sensitive to mislabeling and outliers, as it is based on classical estimators of location and scatter. 
To illustrate this, we consider a bivariate toy example with two classes to which we apply the CQDA classification rule. We generated data from two bivariate normal distributions, depicted in the top left panel of Figure~\ref{fig:noisyQDA}. The dataset contains 80 observations of class 1 (orange) and 100 observations of class 2 (blue). The tolerance ellipses correspond to the points $\bx$ whose Mahalanobis distance 
\begin{equation}
\label{eq:mahalan}
  \text{MD}(\bx,\bar{\bx}_g,\bS_g) = 
  \sqrt{(\bx- \bar{\bx}_g)^\top \bS_g^{-1} 
	(\bx- \bar{\bx}_g)}
\end{equation}
equals {\scriptsize $\sqrt{\chi^2_{2,0.99}}$}\,, the square root of the 0.99 quantile of the $\chi^2$ distribution with $p=2$ degrees of freedom. They visualize the shape of the empirical covariance matrices and fit the data nicely. The grey curve is the quadratic decision boundary obtained by CQDA. We see that it separates the classes quite well, misclassifying only three orange instances into the blue class. This is natural as there is some overlap between the classes. 

In the top right panel we have introduced mislabeling by replacing the label of four observations from each class by the label of the other class. The blown up ellipses and the modified decision boundary show how much CQDA is affected. (The dotted curve is the decision boundary based on the uncontaminated data.) Although the mislabeled cases have no outlying measurements, they are outlying with respect to the class they are (incorrectly) assigned to. This changes the empirical mean and covariance matrix and therefore also the CQDA discriminant scores.   
 
In a second experiment we replaced five observations from class 1 and eight observations from class 2 by outlying points, thereby introducing measurement noise. The blue outliers are positioned such that CQDA based on the clean data would assign them to the orange class. The result is shown in the lower left panel of Figure~\ref{fig:noisyQDA}. The outliers have perturbed the classification, as both the decision boundary and the tolerance ellipses have changed substantially. Because of this the cluster of blue outliers is now classified into the blue class, instead of the orange class in which the uncontaminated boundary would put them.

Finally, the bottom right panel shows the dataset with both mislabeling and outliers, affecting the decision boundary and the ellipses even more.

One must thus be careful when applying CQDA in practice, particularly in situations where mislabeling and/or contamination occur frequently. Note that when the dimension is above 3, the effects in Figure \ref{fig:noisyQDA} are no longer visible by eye.

\subsection{Real-time robust discriminant analysis} 
\label{subsec:RTRQDA}

In order to make DA more reliable in the presence of label and/or measurement noise, several robust alternatives have been proposed. The most common strategy is to replace the classical estimators by robust counterparts. For example, \cite{chork1992integrating} apply the Minimum Volume Ellipsoid estimator introduced in \cite{rousseeuw1984least}, \cite{He:discrim} and \cite{croux2001robust} rely on S-estimators for linear discriminant analysis, whereas \cite{hubert2004fast} propose to use the Minimum Covariance Determinant (MCD) estimator of \cite{rousseeuw1984least}. While this approach yields a more reliable version of DA, all these robust estimators become increasingly computationally demanding at large datasets. Particularly in industrial settings where large amounts of data have to be processed near-instantly, the existing algorithms become infeasible.
 
To address this issue we propose to incorporate the recently introduced real-time deterministic MCD (RT-DetMCD) method of \citep{de2020real}. It is a parallel algorithm for the MCD estimator which runs very fast. For each class $g$ the MCD estimator aims to find the subset of $h_g$ observations whose sample covariance matrix has the lowest determinant. The raw MCD estimates of location and scatter are then computed as the classical mean and covariance matrix of the $h_g$ observations in this subset. The number $h_g$ should be chosen such that $\floor{(n_g+p+1)/2} \leqslant h_g < n_g$ and such that $n_g-h_g$ is above the actual number of cases in class $g$ that are contaminated by label or measurement noise. Since this number is unknown, we take $h_g =\floor{(n_g+p+1)/2}$ to be able to withstand up to 50\% of outliers in each class. Note that $h_g = 0.75\,n_g$ is also often recommended, as it yields more efficient estimates and a degree of robustness that is sufficient in applications with less than 25\% of contamination in each class. 

The MCD estimator can be computed using the FastMCD \citep{rousseeuw1999fast} or DetMCD \citep{hubert2012deterministic} algorithms, but their computational cost is still too high for our purpose. The RT-DetMCD estimator takes a different approach by first splitting up the data into $q$ blocks which are processed in parallel. In each block an improved version of the DetMCD algorithm is run, yielding $q$ estimates of location and scatter. These estimates are then combined through a novel robust pooling strategy which results in a single raw location and scatter estimate. This raw estimate allows a reweighting step in which the classical mean and covariance are computed on all unflagged observations, yielding the final reweighted location and scatter estimate. More details on the algorithm can be found in the Additional Material or \citep{de2020real}.

The RT-DetMCD algorithm is incorporated into QDA as follows. For each class $g=1,\ldots,G$ in the training data we compute the RT-DetMCD estimates of location and scatter, denoted by $\hbmu_{g,\text{R}}$ and $\hbSigma_{g,\text{R}}$ where the subscript $\text{R}$ stands for robust. In order to estimate the class probabilities $p_g$ robustly we use the following procedure \citep{hubert2004fast}: first we compute the robust distance of every observation $\bx_i$ to its own class $y_i$ by
\begin{equation}
	\RD_{i,y_i} = \RD(\bx_i,\hbmu_{y_i,\text{R}} , 
	\hbSigma_{y_i,\text{R}}) = 
	\sqrt{{(\bx_i- \hbmu_{y_i,\text{R}})^\top 
	\hbSigma_{y_i,\text{R}}^{-1} 
	(\bx_i- \hbmu_{y_i,\text{R}})}}\;.
\label{eq:rob_mahalan}
\end{equation}  
Next we flag outliers within each class: we consider $\bx_i$ an outlier when the robust distance with respect to its own class is too large, i.e.\ when $\RD_{i,y_i} >$ {\scriptsize $\sqrt{\chi^2_{p,0.99}}$}\,. Note that these outliers can be the result of label and/or measurement noise. We now drop these outliers from the calculation of the membership probabilities, yielding robust estimates of the membership probabilities  $\hat{p}_{g,\text{R}} = \tilde{n}_g/\tilde{n}$ where $\tilde{n}_g$ denotes the number of non-outliers in class $g$ and $\tilde{n} = \sum_{g=1}^{G}{\tilde{n}_g}$\,. Including this $\hat{p}_{g,\text{R}}$ together with $\hbmu_{g,\text{R}}$ and $\hbSigma_{g,\text{R}}$ into~\eqref{eq:QDA} yields robust discriminant scores.

On top of this we add another feature. Most classifiers have the disadvantage that a new case $\bx$ will always be assigned to one of the known classes. In practice however, it is possible that the new case belongs to a different class which was not present in the training data. We therefore incorporate an anomaly detection step into QDA by assigning a new observation to the `overall outlier' class with label 0 if its robust distance with respect to {\it all} classes is too large. A similar idea was used in the SIMCA method \cite{wold1977simca}. Any unknown observation $\bx$ is thus assigned to the class $g$ for which the discriminant score is highest, under the condition that it does not deviate too much from all known classes. In the latter case, $\bx$ will be given the label $0$.

Putting all of this together, we obtain the proposed real-time robust quadratic discriminant analysis (RT-RQDA) classification given by

\begin{center}
{\bf if}\; $\min_g \RD(\bx, \hbmu_{g,\text{R}},
  \hbSigma_{g,\text{R}}) >$ 
	{\scriptsize $\sqrt{\chi_{p,0.99}^2}$}\; 
	{\bf then}\; assign $\bx$ to class 0;\\ 
{\bf else}\; assign $\bx$ to the class $g$ for which
  $d(\bx, \hbmu_{g,\text{R}}, \hbSigma_{g,\text{R}},
	\hat{p}_{g,R})$ is highest.
\end{center}

\begin{figure}[!ht]
\centering
\includegraphics[width=\linewidth]{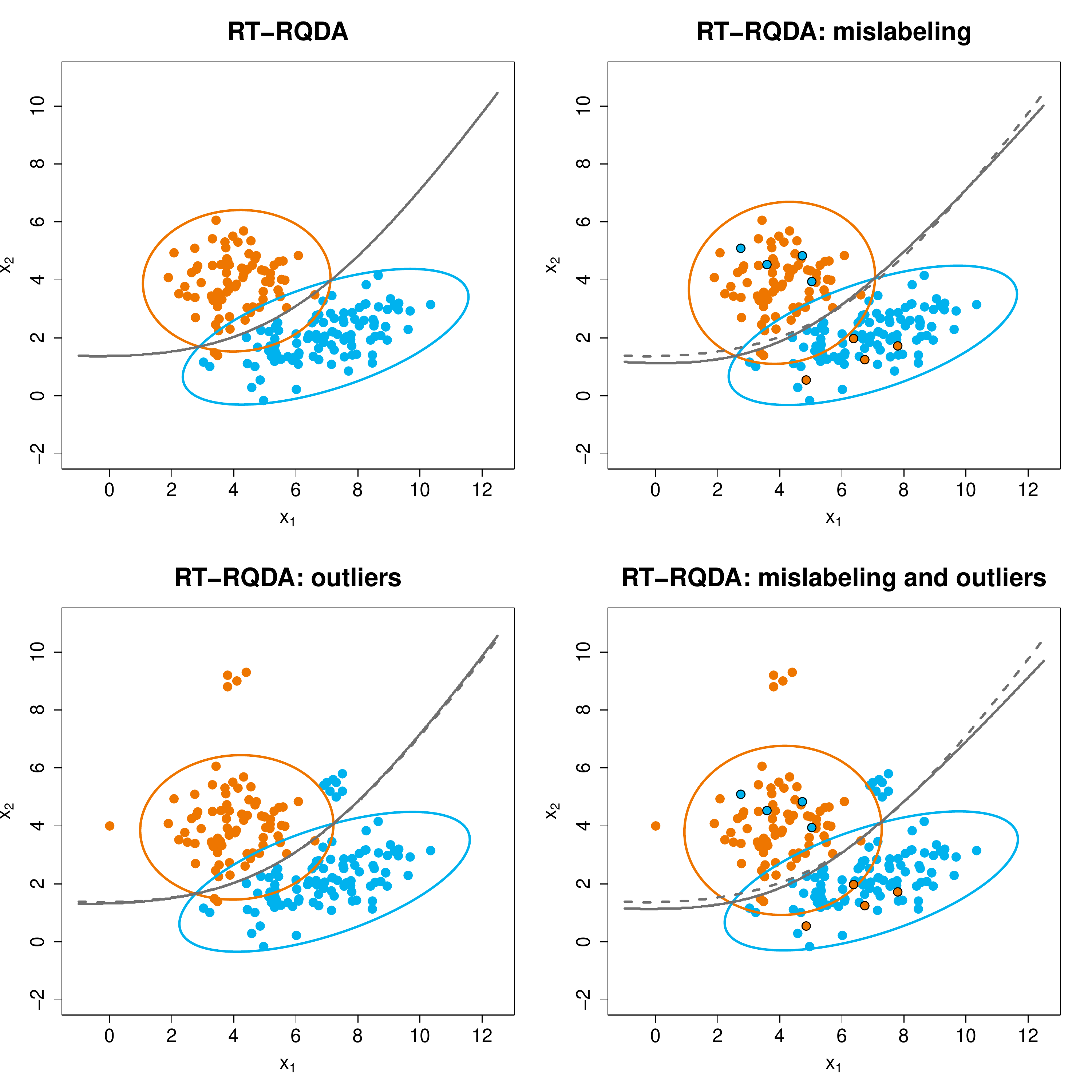}
\caption{The effect of label and measurement noise on robust quadratic discriminant analysis.}
\label{fig:noisyRTRQDA}		
\end{figure}

In order to illustrate this classifier we reconsider the toy example of Subsection~\ref{subsec:CQDA}. Figure~\ref{fig:noisyRTRQDA} shows the result of training RT-RQDA on the clean dataset and on the datasets corrupted with label and/or measurement noise. The top left panel shows that RT-RQDA acts like classical QDA in the absence of noise, as the decision boundary and the tolerance ellipses are almost identical to those in the top left panel of Figure~\ref{fig:noisyQDA}. The three other panels contain the same contamination as in Figure~\ref{fig:noisyQDA}, but RT-RQDA now yields decision boundaries and tolerance ellipses that are very similar to those from the uncontaminated data. This robustness to label and measurement noise will be studied more extensively in the simulation study in Section~\ref{sec:simulation}.

\section{The label bias plot}\label{sec:LBplot}

While the robust discriminant analysis method introduced above performs well under label and measurement noise, by itself it gives little insight into the presence or absence of such noise in individual observations. For that purpose we construct a graphical display which visualizes label and/or measurement noise in the data.

Suppose we have an observation $\bx_i$ with observed class label $y_i$ which is assigned to the class $\hat{y}_i$\,, so the discriminant score $d(\bx_i,\hat{\bmu}_{\hat{y}_i,R}, \hat{\bSigma}_{\hat{y}_i,R}, \hat{p}_{\hat{y}_i,R})$ is the highest. If $\hat{y}_i = y_i$ the QDA classifier has assigned $\bx_i$ to its given class, hence $d(\bx_i,\hat{\bmu}_{\hat{y}_i,R}, \hat{\bSigma}_{\hat{y}_i,R}, \hat{p}_{\hat{y}_i,R}) =  d(\bx_i,\hat{\bmu}_{{y}_i,R}, \hat{\bSigma}_{{y}_i,R}, \hat{p}_{{y}_i,R})$. If on the other hand  $d(\bx_i,\hat{\bmu}_{\hat{y}_i,R}, \hat{\bSigma}_{\hat{y}_i,R}, \hat{p}_{\hat{y}_i,R}) >  d(\bx_i,\hat{\bmu}_{{y}_i,R}, \hat{\bSigma}_{{y}_i,R}, \hat{p}_{{y}_i,R})$ it follows that $\hat{y}_i \neq y_i$\,, so the observation $\bx_i$ was predicted to belong to a class different from its given one. The larger this difference, the more the DA classifier wants to assign $\bx_i$ to its predicted class $\hat{y}_i$ instead of its given class. A high difference could be caused by one or more of the following:
\begin{itemize}
\item the observation $\bx_i$ may have been mislabeled (label noise);
\item the observation $\bx_i$ may be outlying with respect to its given class $y_i$ (measurement noise);
\item there may be overlap between the classes $y_i$ and $\hat{y}_i$ making them difficult to separate.
\end{itemize} 

To quantify label noise we define the \textit{label bias} of the observation $\bx_i$ as
\begin{equation}\label{eq:LB}
\LB(\bx_i) = \sqrt{d(\bx_i,\hat{\bmu}_{\hat{y}_i,R}, \hat{\bSigma}_{\hat{y}_i,R}, \hat{p}_{\hat{y}_i,R}) -  d(\bx_i,\hat{\bmu}_{{y}_i,R}, \hat{\bSigma}_{{y}_i,R}, \hat{p}_{{y}_i,R})}\,.
\end{equation}
The label bias is exactly zero for well-classified cases, and when it is strictly positive the predicted label differs from the given label. A high label bias raises suspicion that the given label might be wrong (mislabeling). 

For quantifying potential measurement noise we use the robust distance~\eqref{eq:rob_mahalan} of $\bx_i$ to its given class $y_i$. This measures how close $\bx_i$ lies to the center of its class, relative to the scatter of its class. As described in Subsection~\ref{subsec:RTRQDA}, high robust distances can be used to flag outliers.

We now introduce the \textit{label bias plot} (LB-plot) of class $g$ as the scatter plot of the points
\begin{equation} \label{eq:LBplot}
\left(\text{RD}(\bx_i, \hat{\bmu}_{y_i,R}, \hat{\bSigma}_{y_i,R}), \LB(\bx_i)\right)
\end{equation}
for all $\bx_i$ that belong to same class $g$, i.e.\ with $y_i=g$. Note that we defined LB in~\eqref{eq:LB} as a square root, because the discriminant scores contain the squared robust distances and we want to put the axes of the label bias plot on the same footing.

In the label bias plot the points are colored according to their predicted class label $\hat{y}_i$. As one plot is made for each class, we obtain $G$ different LB-plots. High values of LB suggest mislabeling, whereas points with high RD are outlying with respect to their observed class.  

\begin{figure}[!ht]
\centering
\includegraphics[width=\linewidth]{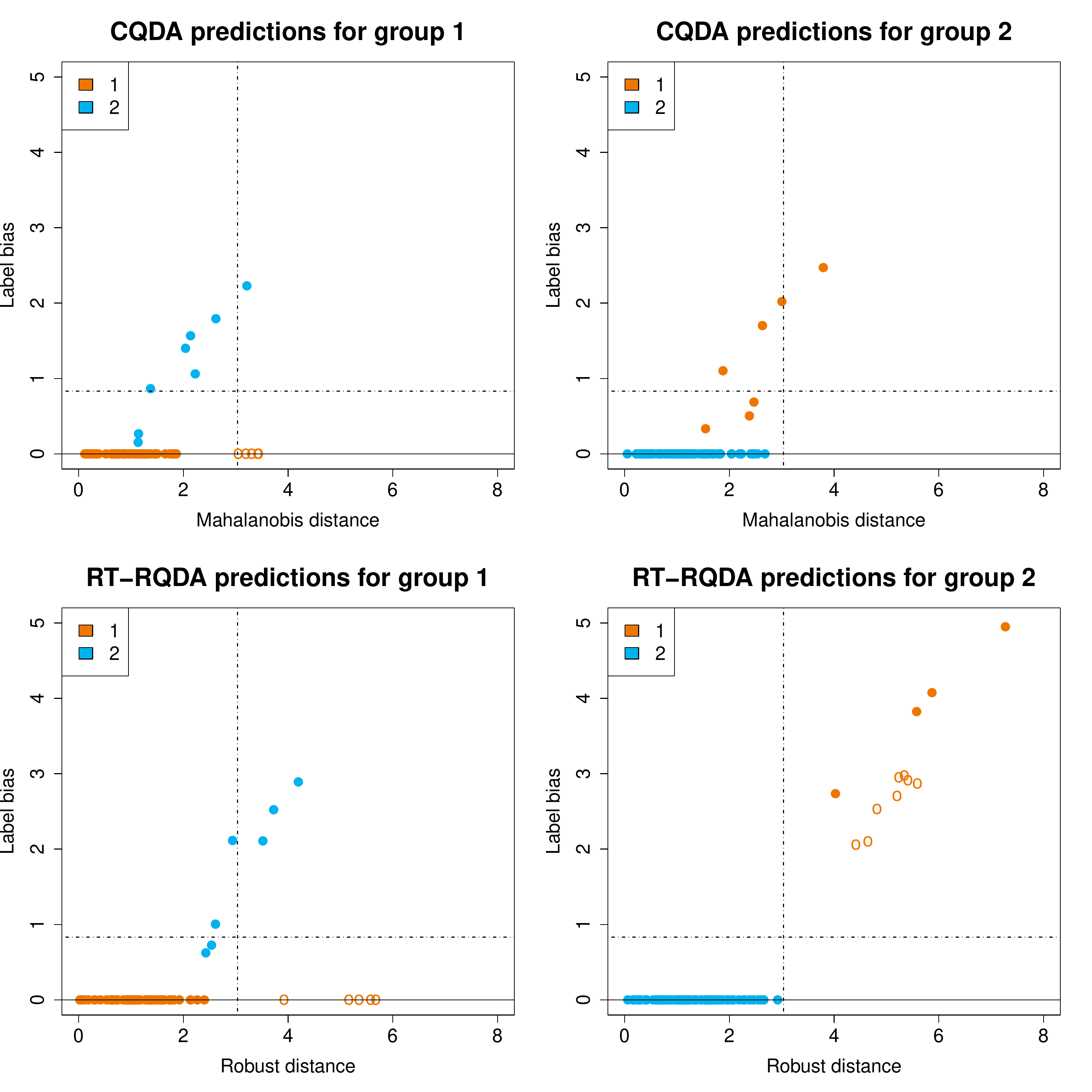}
\caption{LB-plots of classical and robust QDA applied to artificial data with label and measurement noise.}
\label{fig:LBplots}		
\end{figure}

Figure~\ref{fig:LBplots} presents the LB-plots of the artificial data with mislabeling and outliers shown in the lower right panel of Figure~\ref{fig:noisyRTRQDA}. The top panel shows the LB-plots resulting from CQDA, whereas the bottom panel shows those for RT-RQDA.

Note that each LB-plot has two dashed lines, corresponding to cutoffs on the axes. For the robust distances on the horizontal axis we use the cutoff value c={\scriptsize $\sqrt{\chi^2_{p, 0.99}}$} that we already used in Subsection~\ref{subsec:RTRQDA} to flag outliers in a class. For the LB on the vertical axis we use the cutoff \small{$\sqrt{\ln(2)}=0.83$}. Points above this cutoff have a mixture model likelihood $p_gf_g(\bx_i)$ that is at least twice as high for their predicted class $\hat{y}_i$ than for their given class $y_i$\,.

Finally, we depict the overall outliers, i.e. the observations for which\linebreak $\min_g \RD(\bx_i, \hat{\bmu}_{g, R}, \hat{\bSigma}_{g, R}) > c$, as empty circles rather than dots. These observations are outlying with respect to every class in the data, suggesting they may be gross errors or members of a different population. 

We first discuss the LB-plots of RT-RQDA in the lower half of Figure~\ref{fig:LBplots}, since they are the most informative. Looking at the LB-plot for class 1 (orange), we immediately note that most points have $\LB=0$ and lie to the left of $c$. These are the regular observations from class 1 that are classified to belong to class 1. Next, we see five points with $\LB = 0$ and a robust distance above the cutoff $c$. They are displayed as empty circles, indicating that they are also outlying with respect to class 2. They correspond to the orange outliers in Figure~\ref{fig:noisyRTRQDA}, which were generated to be closer to class 1 than to class 2. Hence they are not misclassified but the LB-plot indicates that they do not sit well in the orange class. The plot also contains seven points with a strictly positive label bias, meaning they have been assigned to the blue class although their given label is orange. Three of them have a relatively low LB, and a low robust distance indicating that they are within the robust tolerance ellipse of class 1. The LB-plot thus tells us that while the classifier assigns these points to class 2, they are also quite close to class 1. This suggests overlap, and indeed in Figure~\ref{fig:noisyRTRQDA} these three points lie in the region where the ellipses overlap. There are also three points of class 1 with a high LB and $\RD > c$. This suggests label or measurement noise, and these three points were indeed deliberately mislabeled in Figure~\ref{fig:noisyRTRQDA}. Note that the LB-plot does not allow us to determine the root cause of their outlyingness: it could be that the observations belong to class 1 but due to measurement noise they ended up closer to class 2, or they could actually belong to class 2 but received label 1 due to label noise. In either case the points would show up in the upper right quadrant of the LB-plot. Since the points are plotted as dots and not as empty circles, they are not outlying with respect to class 2. This lends more credence to label noise than measurement noise. Finally there is one point with $\LB \approx 2$ and a robust distance which is slightly smaller than $c$. This is a borderline case, which could be due to overlap or mislabeling. On the scatter plot in Figure~\ref{fig:noisyRTRQDA} we see that one of the mislabeled points of class 1 indeed belongs to the overlapping region.
 
The LB-plot of class 2 indicates that most points are well classified, but there are 12 points with large LB and large RD. Four of them are plotted as dots, indicating that they do not have a large robust distance with respect to class 1. These correspond to the mislabeled points in Figure~\ref{fig:noisyRTRQDA}. The other eight points in the top right of the LB-plot are displayed as empty circles and are therefore outlying with respect to all classes. They correspond to the small blue cluster of outliers in Figure~\ref{fig:noisyRTRQDA}, which indeed lies closer to the orange class yet it is also outlying to that class.

The LB-plots based on RT-RQDA thus gave us insight into how well the cases were classified, whether there is overlap between the classes, and whether there are any suspicious labels or outliers. This information becomes even more useful in higher dimensions, when there is no version of Figure~\ref{fig:noisyRTRQDA} to look at.

The LB-plots based on CQDA are shown in the top panels of Figure~\ref{fig:LBplots}. The horizontal axis now shows the classical Mahalanobis distances~\eqref{eq:mahalan} based on the sample mean and covariance matrix. The label bias is defined as in~\eqref{eq:LB} but now based on the classical discriminant scores. Although we see some similarities with the LB-plots from RT-RGDA, there are several important differences. For class 1, only four out of the five overall outliers are flagged with empty circles, and only one of the mislabeled points has a distance above $c$. The other mislabeled points look less suspicious, instead the plot rather suggests they are in an overlapping region. In the LB-plot of class 2 only one point has a distance above $c$. Indeed, in the bottom right panel of Figure~\ref{fig:noisyQDA} the CQDA tolerance ellipses were so inflated that they engulf all blue points but one. In general, the label and measurement noise is less visible in the LB-plots based on CQDA than in those based on RT-RQDA. The LB-plots of CQDA would lead us to think that the misclassifications are mainly due to a large overlap between the classes, which we know is not the ground truth in these generated data. In practice we therefore recommend to use the LB-plots based on the robust estimates whenever available.

\section{Simulation study} \label{sec:simulation}
In this section we compare the performance of the proposed RT-RQDA classifier to that of classical QDA through a simulation study. We will denote the true classes as $\pi_g$ for $g = 1,\ldots, G$.

\subsection{Simulation setup}
We first generate uncontaminated (clean) data $\bX$ with label vector $\by$. We create three classes $\pi_1$, $\pi_2$ and $\pi_3$ of $p=5$ dimensional observations. The observations of class $\pi_g$ are sampled from a normal distribution $N(\bmu_g,\bSigma_g)$, where the class centers are given by $\bmu_1 = (6,0,0,0,0)^\top$, $\bmu_2 = (0,0,6,0,0)^\top$ and $\bmu_3 = (0,0,0,0,6)^\top$. The covariance matrices of the classes are the diagonal matrices $\bSigma_1 = I_5$, $\bSigma_2 =\diag(1,2,3,4,5)$ and $\bSigma_3 =\diag(1,1,1,5,10)$. The number of observations in each class is set to $n_1 = 250,000$, $n_2 = 350,000$ and $n_3 = 400,000$ with $n=1,000,000$ to emulate industrial data sizes. The true prior probabilities of the classes $\pi_1$, $\pi_2$ and $\pi_3$ are thus 0.25, 0.35 and 0.4\,.

In order to evaluate the performance of the method in more realistic scenarios, we also consider three simulation setups with contamination. In the first we only include label noise. In this setting the data is corrupted according to the {\it noisy completely at random} (NCAR) process~\cite{frenay2013classification}. For each class $\pi_g$ we randomly relabel $\varepsilon_\ell n_g$ observations: $\varepsilon_\ell n_g/2$ instances receive one of the other labels $k \ne g$, while the other half acquire the remaining label. We choose the label noise fraction $\varepsilon_\ell = 20\%$.  

The second setup contains measurement noise but no label noise. The measurement noise is generated as follows. For class $\pi_1$ the outliers are sampled from cluster contamination at $N(\bmu_1^*,\bSigma_1^*)$ with $\bmu_{1}^* = (-6,0,0,0,0)^\top$ and $\bSigma_{1}^* = \frac{1}{10} \bSigma_1$. For class $\pi_2$ we consider point contamination by concentrating all outliers into a single point $\bmu_2^*=(0, 0, -15, 0, 20)^\top$. For the third class $\pi_3$ we generate shift contamination according to $N(\bmu_3^*,\bSigma_3)$ with $\bmu_{3}^* = (14,0,0,0,-6)^\top$. We replace $\varepsilon_m n_g$ random observations of each class $g$ by outliers of the corresponding types, where $\varepsilon_m$ denotes the fraction of measurement noise, which we set to $20\%$ in this simulation. Note that the contamination has been generated in such a way that the contaminating points are outlying with respect to all classes. Hence, we expect a good classifier to predict them as `overall outliers', that is, assign them to the additional class $\pi_0$.
  
In the final setup we combine both types of noise by introducing $\varepsilon_\ell/2$ label noise and $\varepsilon_m/2$ measurement noise, both as described above.

\subsection{Evaluation of results}

Each simulation setup is replicated 50 times and the performance of  RT-RQDA is compared to CQDA in several ways. First of all, we compare the misclassification errors of both methods by computing the average confusion matrix over all replications. The standard confusion matrix will not allow us to properly analyze the results under label and measurement noise. Therefore, we will present extended confusion matrices, for which we need some notation. Consider class $\pi_g$\,. The observations in this class can be split up in up to 4 subclasses, depending on the contamination scheme. The largest subclass is that of the clean observations, which were generated from $N(\bmu_{g}, \bSigma_{g})$ and received the correct label. This subclass is denoted $\pi_{g,g}$\,. Secondly, there are the observations which were generated from the clean distribution of class $\pi_g$ but which received a wrong label. We denote these as $\pi_{g,k}$ and $\pi_{g,\ell}$ where $k$ and $\ell$ are the labels of the other two classes. Finally, there are the observations with measurement error denoted as $\pi_{g,0}$. 

Instead of creating confusion matrices with three rows, one for each class, we split up each class into its nonempty subclasses. For the experiment with clean data, we still have only three rows. The experiment with only label noise has three subclasses in each class, and hence yields an extended confusion matrix with nine rows. When only measurement noise occurs each class has two subclasses, yielding an extended confusion matrix with six rows. The experiment with both label and measurement noise has all twelve rows. By extending the confusion matrices in this way, we obtain very precise insight into what happens with every type of data point, corrupted or otherwise. Finally, we also add an additional column to the confusion matrix, denoted $\pi_0$, which shows the percentage of points classified as an overall outlier using the outlier detection rule of RT-RQDA.

In addition to the extended confusion matrices we also report the average Kullback–Leibler (KL) divergence. The KL divergence of the estimated scatter matrix $\hbSigma_g$ from the true covariance matrix $\bSigma_g$ of class $\pi_g$ is defined as
\begin{equation}
\mbox{KL}_g = \mathrm{trace}(\hbSigma_g \bSigma_g^{-1})
  - p - \ln|\hbSigma_g \bSigma_g^{-1}|.
\end{equation}
The determinants of the true covariance matrices are $|\bSigma_1| = 1$, $|\bSigma_2| = 120$ and $|\bSigma_3| = 50$. The estimated determinants $|\hbSigma_g|$ are also reported. Ideally, the estimated determinants should be close to the theoretical ones as this suggests that they accurately describe the volume of the clean data cloud. Finally, for each class we also report how many observations are flagged as outlying with respect to their own class. These are the observations $\bx_i$ for which $\RD_{i,y_i} > c$ where $\RD_{i,y_i}$ is the robust distance~\eqref{eq:rob_mahalan}. We normalize this number by the total number of noisy observations in class $\pi_g$ given by $(\varepsilon_\ell + \varepsilon_m) n_g$\,, and denote the resulting quantity by $\alpha_g$.  Ideally, $\alpha_g$ should be close to one, as this indicates the label and measurement noise are well detected. Our simulations were done in MATLAB on an Intel Core i7-8700K processor based computer with 16 GB of 3.70GHz RAM. The same hardware was used in all experiments. 

\subsection{Results on clean data}
We first consider the setting of clean data, for which the results are presented in Table~\ref{tbl:sim_nocntm}. The confusion matrix has only three rows, one for each class, since no label noise or measurement noise was introduced. Since we have set the outlier cutoff to {\scriptsize $\sqrt{\chi^2_{p,0.99}}$}\,, perfect classification corresponds with diagonal values equal to 0.99 and a $\pi_0$ column equal to 1\%. We see that both classification methods perform very well on clean data. Note that CQDA is somewhat more efficient in this clean setting as it has lower KL values than RT-RQDA, and the determinants of the estimated covariance matrices are closer to the true values. The tolerance ellipsoids of RT-RQDA have somewhat lower volumes, resulting in slightly higher values in the column of $\pi_0$.

\begin{table}[!ht]			
\centering
\caption{Classical and robust discriminant analysis results for uncontaminated data.}
\label{tbl:sim_nocntm}
\vspace{0.3cm}
\resizebox{0.8\columnwidth}{!}{ % was 1\columnwidth
\begin{tabular}{@{}lrrrrrrrrrrr@{}}
\toprule[1.5pt]	
	& \multicolumn{4}{c}{CQDA} && \multicolumn{4}{c}{RT-RQDA} \\
  & \multicolumn{1}{r}{$\pi_1$}	& \multicolumn{1}{r}{$\pi_2$} & \multicolumn{1}{r}{$\pi_3$} & \multicolumn{1}{r}{$\pi_0$} && \multicolumn{1}{r}{$\pi_1$} &  \multicolumn{1}{r}{$\pi_2$} & \multicolumn{1}{r}{$\pi_3$} & \multicolumn{1}{r}{$\pi_0$} \\ 
			\cmidrule{2-5} \cmidrule{7-10} 	
\multicolumn{9}{@{}l}{\textbf{Extended confusion matrices}}\\			
			{$\pi_{1, 1}$}	&         0.990 &   0.000  &  \phantom{f}0.000  &  \phantom{ff}0.010  & \hspace{1em} &  0.986  & \phantom{f}0.000 &  \phantom{ff}0.000 &  \phantom{ff}0.013	\\
			{$\pi_{2, 2}$}	&         0.000 &   0.981  &  0.010  &  0.009  &&  0.000  & 0.978 &  0.009 &   0.013	\\
			{$\pi_{3, 3}$}	&         0.000 &   0.004  &  0.986  &  0.009  &&  0.000  & 0.005 &  0.983 &   0.013	\\ 
			\; \\
\multicolumn{9}{@{}l}{\textbf{Performance metrics}}\\	
			{KL}	 		&     $<$0.001  &   $<$0.001  & $<$0.001   & && 0.007  &  0.007  &  0.007  	\\	
			{$|\hbSigma|$}&     1.001  & 120.100  & 49.994  & && 0.764  &  91.750 & 38.195 	\\	
\bottomrule
\end{tabular}
}
\end{table}

\subsection{Results under label noise}

The results of the simulation setup with $20\%$ label noise are summarized in Table~\ref{tbl:sim_labelcntm}. For CQDA the performance metrics show large increases in the KL divergences and in the determinants of the estimated covariance matrices. Although CQDA fails to accurately estimate the model parameters, its classification error remains rather low as can be inferred from its extended confusion matrix. This is mainly due to the fact that the true classes are well separated. Unlike CQDA, RT-RQDA shows a very stable behavior, with virtually unchanged estimates of the model parameters and a confusion matrix that is similar to the clean setting. Finally, note that the $\alpha_g$ values suggest that CQDA identifies only a small fraction of the mislabeled observations as outliers with respect to the class of their given label, whereas RT-RQDA flags nearly all of them.

\begin{table}[!ht]			
\centering
\caption{Classical and robust discriminant analysis results with $20\%$ label noise.}
\label{tbl:sim_labelcntm}
\vspace{0.3cm}
\resizebox{0.8\columnwidth}{!}{
		\begin{tabular}{@{}lrrrrrrrrrr@{}} 
			\toprule[1.5pt]	
			& \multicolumn{4}{c}{CQDA} && \multicolumn{4}{c}{RT-RQDA} \\
			& \multicolumn{1}{r}{$\pi_1$}	& \multicolumn{1}{r}{$\pi_2$} & \multicolumn{1}{r}{$\pi_3$} & \multicolumn{1}{r}{$\pi_0$} && \multicolumn{1}{r}{$\pi_1$} &  \multicolumn{1}{r}{$\pi_2$} & \multicolumn{1}{r}{$\pi_3$} & \multicolumn{1}{r}{$\pi_0$} \\ 
			\cmidrule{2-5} \cmidrule{7-10} 
\multicolumn{9}{@{}l}{\textbf{Extended confusion matrices}}\\	
			{$\pi_{1, 1}$}	&   0.999  & 0.000 &  \phantom{f}0.000   &  \phantom{fff}0.000 & \phantom{ff}&    0.986  & \phantom{fff}0.000  & \phantom{fff} 0.000   &  \phantom{ffff}0.013			\\
			{$\pi_{1, 2}$}	&   0.999  & 0.000 &  0.000   &  0.000 &&     0.986  & 0.000  &  0.000  &  0.013			\\
			{$\pi_{1, 3}$}	&   0.999  & 0.000 &  0.000   &  0.000 &&     0.986  & 0.000  & 0.000    & 0.013			\\
			{$\pi_{2, 1}$}	&   0.001  & 0.975  &   0.021  &  0.004  && 0.000 &   0.979 &   0.008  &   0.012			\\
			{$\pi_{2, 2}$}	&   0.001  & 0.975  &   0.020  &  0.004  && 0.000 &   0.979 &   0.008  &   0.012			\\
			{$\pi_{2, 3}$}	&   0.001  & 0.975  &   0.021  &   0.004 && 0.000 &   0.979  &   0.008   & 0.012			\\
			{$\pi_{3, 1}$}	&   0.003  &  0.002 &    0.992  &  0.003  && 0.000   & 0.005   &   0.982    & 0.012			\\
			{$\pi_{3, 2}$}	&   0.003  &  0.002 &    0.992  &  0.003  && 0.000   & 0.005   &   0.982    & 0.013			\\
			{$\pi_{3, 3}$}	&   0.003  &  0.002 &    0.992  &  0.003  && 0.000   & 0.005   &   0.982    & 0.013			\\
			\; \\
\multicolumn{9}{@{}l}{\textbf{Performance metrics}}\\	
			{KL}	 		&     11.123  &    2.394   &    2.315 &  	  &&   0.007   &  0.007   & 0.007		  	\\	
			{$|\hbSigma|$}&     143.230 &    1047.600   &    508.410 &  &&   0.766   &  97.165   &    38.609			\\	
			{$\alpha$} 	&	  0.329   &   0.133   &   0.383 &   	  &&   0.949   &  0.926   &    1.082			\\ 
			\bottomrule
		\end{tabular}
	}
\end{table}

\subsection{Results under measurement noise}
Table~\ref{tbl:sim_measurementcntm} reports the results when the data contains 20\% of measurement noise. Since by construction all contaminated points are outlying to all classes, they should be classified into $\pi_0$. The extended confusion matrix now has six rows because each class has a subclass generated as outliers ($\pi_{g,0}$) and a subclass generated as clean data ($\pi_{g,g}$). We see major differences between CQDA and RT-RQDA. First note that CQDA barely detects outliers as indicated by the low values in the column of $\pi_0$. Instead, it classifies the outliers as regular observations. The performance metrics in the table reveal the problem with CQDA: the estimates of the parameters are affected so strongly by the outliers that the resulting model does not detect them anymore. This illustrates the so-called masking effect \citep{maronna2019robust}. In contrast, RT-RQDA correctly detects the introduced noise as outlying observations. A second important effect is that with CQDA the classification of the clean data suffers. In particular, the regular observations of class $\pi_1$ and class $\pi_3$ have a misclassification rate of roughly 20\% and 40\%. The proposed RT-RQDA method instead has a stable performance which is comparable to that in the clean scenario of 
Table~\ref{tbl:sim_nocntm}.

\begin{table}[!ht]			
\centering
\caption{Classical and robust discriminant analysis results with $20\%$ measurement noise.}
\label{tbl:sim_measurementcntm}
\vspace{0.3cm}
\resizebox{0.8\columnwidth}{!}{
\begin{tabular}{@{}lrrrrrrrrrr@{}} 
\toprule[1.5pt]	
	& \multicolumn{4}{c}{CQDA} && \multicolumn{4}{c}{RT-RQDA} \\
	& \multicolumn{1}{r}{$\pi_1$}	& \multicolumn{1}{r}{$\pi_2$} & \multicolumn{1}{r}{$\pi_3$} & \multicolumn{1}{r}{$\pi_0$} && \multicolumn{1}{r}{$\pi_1$} &  \multicolumn{1}{r}{$\pi_2$} & \multicolumn{1}{r}{$\pi_3$} & \multicolumn{1}{r}{$\pi_0$} \\ 
	\cmidrule{2-5} \cmidrule{7-10} 	
\multicolumn{9}{@{}l}{\textbf{Extended confusion matrices}}\\	
			{$\pi_{1, 0}$}	&      1.000   &   0.000  &  0.000 &  \phantom{fff} 0.000   &\phantom{ff}&  0.000   &  \phantom{f} 0.000    & \phantom{ff} 0.000   &  \phantom{fff} 1.000	\\
			{$\pi_{1, 1}$}	&      0.794   &   0.000  &  0.205 &  0.000   &&  0.989   &  0.000    & 0.000   &  0.011		\\
			{$\pi_{2, 0}$}	&      0.000   &   1.000  &  0.000 &  0.000   &&  0.000   &  0.000    & 0.000   &  1.000	\\
			{$\pi_{2, 2}$}	&      0.000   &   0.977  &  0.005 &  0.018   &&  0.000   &  0.980    & 0.009   &  0.010		\\
			{$\pi_{3, 0}$}	&      0.193   &   0.000  &  0.782 &  0.026   &&  0.000	  &  0.000    & 0.000   &  1.000	\\
			{$\pi_{3, 3}$}	&      0.193   &   0.191  &  0.614 &  0.003   &&  0.000   &  0.004    & 0.985   &  0.010	\\
			\; \\
\multicolumn{9}{@{}l}{\textbf{Performance metrics}}\\	
	{KL} & 42.500 & 47.844 & 37.411	& && 0.001 & 0.001 & 0.001 & \\	
	{$|\hbSigma|$} & 18.029 & 2180.300 & 1732.700 & && 0.905 & 108.720 & 45.223	& \\
	{$\alpha$} & 0.065 & 0.074 & 0.046 & && 1.046 & 1.045 & 1.045	& \\
\bottomrule
\end{tabular}
}
\end{table}

\subsection{Results under label and measurement noise}

Finally, Table~\ref{tbl:sim_allcntm} presents the simulation results with label and measurement noise occurring simultaneously. Also here RT-RQDA exhibits a stable behavior. All of the far outliers are detected and allocated to the outlier class $\pi_0$. The misclassification errors are similar to those for clean data and the model parameters are estimated accurately, as indicated by the low values of the KL divergence and the determinants of the estimated covariance matrices. 

In contrast, CQDA is heavily affected by the noise. It fails to detect the outliers as can be seen from the low percentages in the column of class $\pi_0$ and the low values of $\alpha_g$. The classification of the clean data is also affected, in particular for subclass $\pi_{3,3}$ which has a misclassification error of 18\%. The estimated parameters deviate substantially from their true values as evidenced by the elevated KL divergences and high determinants. 

From the simulation study we conclude that the proposed RT-RQDA method is more reliable than CQDA in the presence of label and/or measurement noise while maintaining a competitive performance on clean data.

\begin{table}[!ht]			
\centering
\caption{Classical and robust discriminant analysis results with $10\%$ label and $10\%$ measurement noise.}
\label{tbl:sim_allcntm}
\vspace{0.3cm}
\resizebox{0.8\columnwidth}{!}{
\begin{tabular}{@{}lrrrrrrrrrr@{}} 
\toprule[1.5pt]	
	& \multicolumn{4}{c}{CQDA} && \multicolumn{4}{c}{RT-RQDA} \\
	& \multicolumn{1}{r}{$\pi_1$}	& \multicolumn{1}{r}{$\pi_2$} & \multicolumn{1}{r}{$\pi_3$} & \multicolumn{1}{r}{$\pi_0$} && \multicolumn{1}{r}{$\pi_1$} &  \multicolumn{1}{r}{$\pi_2$} & \multicolumn{1}{r}{$\pi_3$} & \multicolumn{1}{r}{$\pi_0$} \\ 
			\cmidrule{2-5} \cmidrule{7-10} 
\multicolumn{9}{@{}l}{\textbf{Extended confusion matrices}}\\	
			{$\pi_{1, 0}$}	&       0.000  & 0.000   &  1.000  &  \phantom{fff} 0.000  &\phantom{ff} &    0.000  &  0.000  & \phantom{ff} 0.000   & \phantom{ff} 1.000	\\
			{$\pi_{1, 1}$}	&       0.940  & 0.000   &  0.060  &  0.000  &&    0.988  &  0.000  & 0.000   &  0.012	\\
			{$\pi_{1, 2}$}	&       0.940  & 0.000   &  0.060  &  0.000  &&    0.988  &  0.000  & 0.000   &  0.012		\\
			{$\pi_{1, 3}$}	&       0.940  & 0.000   &  0.060  &  0.000  &&    0.987  &  0.000  &  0.000   &  0.012		\\
			{$\pi_{2, 0}$}	&       0.000  & 1.000   &  0.000  &  0.000  &&    0.000    &  0.000  &  0.000   &  1.000	\\
			{$\pi_{2, 1}$}	&       0.004  & 0.980   &  0.011  &  0.005  &&    0.000 &     0.981  &  0.009  &  0.011	\\
			{$\pi_{2, 2}$}	&       0.004  & 0.980   &  0.011  &  0.005  &&    0.000  &    0.980  &  0.009   &  0.011	\\
			{$\pi_{2, 3}$}	&       0.004  & 0.980   &  0.011  &  0.005  &&    0.000  &    0.980  &  0.009   &  0.011	\\
			{$\pi_{3, 0}$}	&       0.491  & 0.000   &  0.373  &  0.136  &&    0.000  &    0.000    &  0.000   &  1.000	\\
			{$\pi_{3, 1}$}	&       0.110  & 0.065   &  0.823  &  0.002  && 0.000  &  0.005  &    0.984  &  0.011	\\
			{$\pi_{3, 2}$}	&       0.110  & 0.065   &  0.822  &  0.002  &&  0.000 &  0.005  &    0.983  &  0.011	\\
			{$\pi_{3, 3}$}	&       0.110  & 0.065   &  0.822  &  0.002  && 0.000  &  0.005  &    0.983  &  0.011	\\
			\; \\
\multicolumn{9}{@{}l}{\textbf{Performance metrics}}\\	
			{KL}	 		&     24.547   &    21.399   &    17.698   &&& 0.003  &  0.004   & 0.003 	\\
			{$|\hbSigma|$}&    141.900   &    4596.400 &    2259.200 &&& 0.837  &  106.090   &  42.160 	\\
			{$\alpha$} 	&	  0.251    &   0.159     &    0.246    &&& 1.014  &  0.987   &    1.058 	\\ 
\bottomrule
\end{tabular}
}
\end{table}

\section{Applications to real data}
\label{sec:experiments}

\subsection{Diabetes data}
\label{Pima}
We start by analyzing a benchmark data set which arose out of a study about the onset of diabetes mellitus \cite{Smith1988}. The data were downloaded from \cite{mlbench2015}. The subjects (cases) are females from a community of Pima Native Americans, with classes `positive' (diabetic) and `negative'. For each subject eight variables were recorded: number of pregnancies, plasma glucose concentration, diastolic blood pressure, triceps skin fold thickness, serum insulin, a body mass index (BMI), diabetes pedigree function, and age. Following \cite{Ripley1996} we removed the cases with physically impossible measurements (such as zero blood pressure), leaving us with 392 subjects. Since several variables were skewed, we robustly transformed all variables toward central normality by the technique of \cite{Raymaekers:TVCN} with its default options. We also verified that each of the eight variables has a positive relation to diabetes, in the sense that higher values increase the odds of the disease.

\begin{figure}[!ht]
\centering
\vspace{0.2cm}
\includegraphics[width = 0.99\textwidth]
  {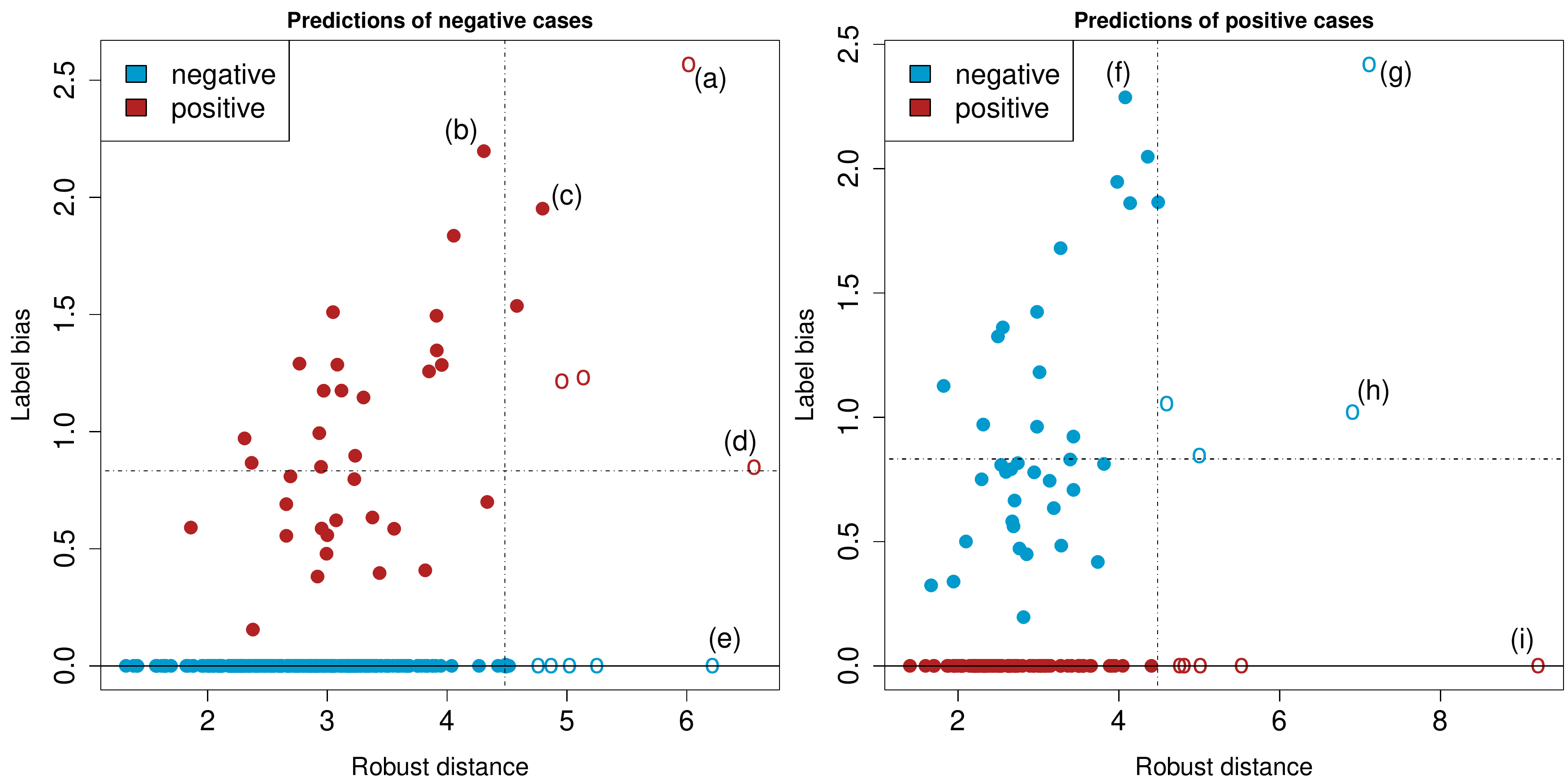}
\caption{LB-plots of the classes of negative (left)
  and positive (right) subjects in the diabetes 
	data, using RT-RQDA.}
\label{fig:Pima_both}
\end{figure}

Applying RT-RQDA yielded the LB-plots in Figure \ref{fig:Pima_both}. The left panel shows the negative class. The correctly classified cases are shown as blue points with label bias LB~$=0$. The red points were misclassified into the positive class. Some of them have a high label bias above the dashed horizontal line, meaning that \mbox{RT-RQDA} considers it more than twice as likely for them to be diabetic rather than negative. The most extreme ones are marked (a), (b) and (c). Case (a) has high glucose, insulin and age, while (b) has high blood pressure and pedigree. Case (c) has high glucose, pedigree, and the highest insulin of the entire dataset. All of these characteristics point to diabetes, arousing suspicion that these subjects were misclassified or have a high probability of being diagnosed in the future.
We also see two points with high robust distance but lower LB. Subject (d) has high blood pressure and very high BMI. Subject (e) is correctly assigned to the negative class since LB~$=0$, but has a high distance partly due to an extremely low blood pressure of 30. Note that (a), (d) and (e) have hollow plot symbols, indicating that their distance to the other class is above the distance cutoff as well, so they are not close to either class. On the other hand, (c) has a high distance from the negative class but its solid plot symbol indicates that it is not so far from the other class.

The LB-plot of the subjects labeled positive for diabetes is in the right panel of Figure \ref{fig:Pima_both}. The subjects with highest label bias are marked (f) and (g). Subject (f) has very low triceps, insulin and BMI, which is in line with the negative class. Subject (g) has extremely low blood pressure and rather low glucose and insulin. These would point to the negative class, hence the high LB. The high BMI of (g) correctly suggests the positive class but is unusual, which increases the robust distance on the horizontal axis. Point (h) corresponds to the highest blood pressure in the entire dataset as well as the highest BMI. Subject (i) has the highest triceps in the dataset as well as the highest pedigree. From the plot symbols of (g), (h) and (i) we conclude that they lie far away from the centers of both classes.

\subsection{Fruit data}
\label{sec:fruit}
We analyze the fruit dataset originally collected by Colin Greensill (Central Queensland University, Rockhampton, Australia), parts of which were analyzed in \cite{hubert2004fast, branden2005robust}. The original dataset contains the result of a spectroscopy experiment conducted on $n=2818$ cantaloupe melons of six different cultivars. Each of the spectra was measured on $256$ wavelengths. We selected four of the six cultivars (labeled ``D'', ``H'',  ``Ha'' and ``E'') for illustrating our methods. The motivation for this selection is twofold. First, the largest cultivar (named ``ES'') was not included because it did not have a roughly elliptical distribution, which makes its classification by QDA less appropriate. Second, the smallest cultivar (named ``M'') was not included because we do not have additional information on it, making the interpretation of the results more difficult. The selected four cultivars yield a dataset of size $1774 \times 256$ which consists of four classes of sizes 490, 180, 500 and 988. It is known that the data from cultivar Ha was collected using different illumination setups and so one might expect subgroups within this class. The previous analyses of \cite{hubert2004fast, branden2005robust} did in fact identify one subgroup within this cultivar corresponding with a different illumination setup.

\begin{figure}[!ht]
	\centering
	\includegraphics[width = 0.8\textwidth]{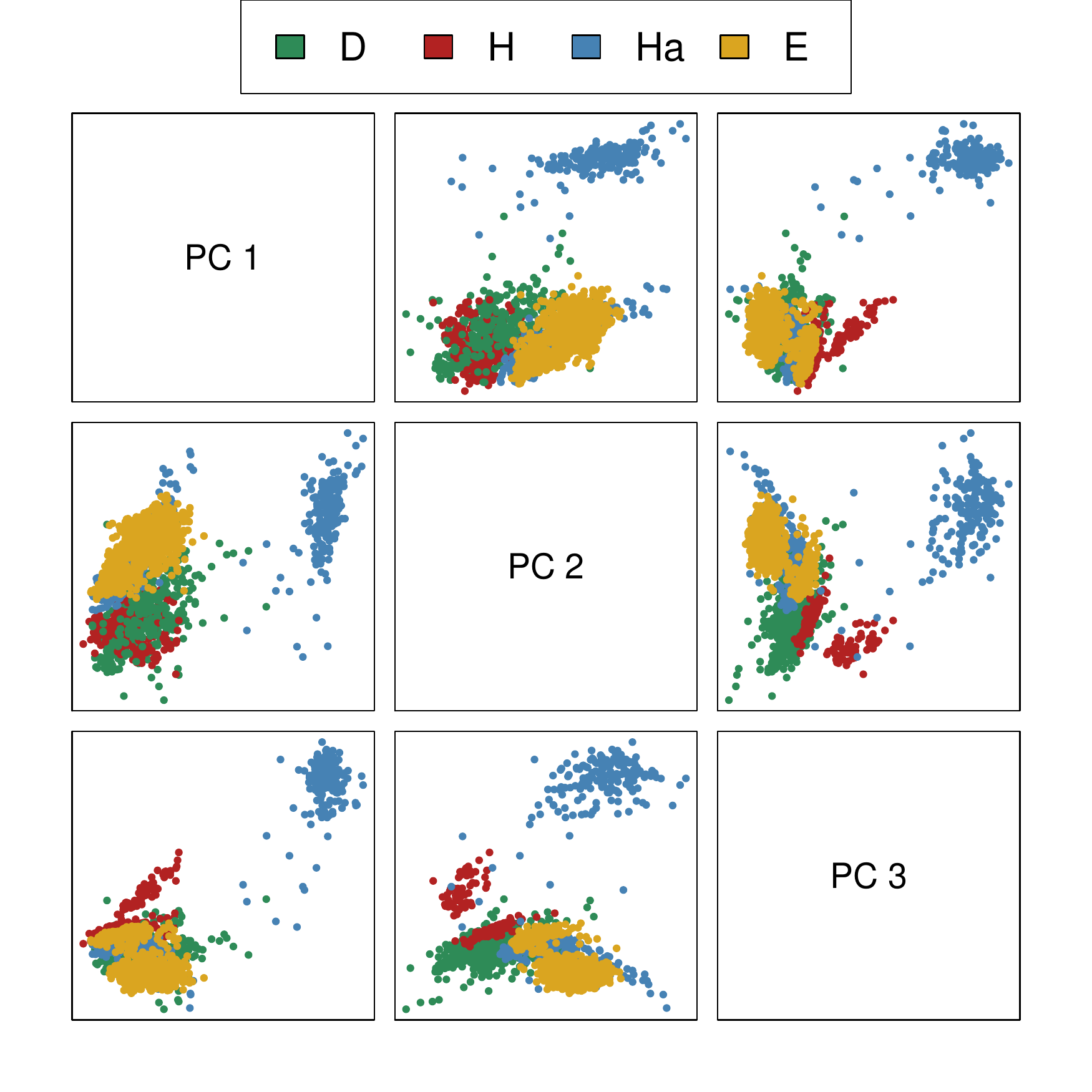}
	\caption{Pairwise scatter plot of the scores of the fruit data after robust principal component analysis.}
	\label{fig:fruit_pca}
\end{figure}

We aim to classify the spectra into the four classes corresponding to the cultivars. It is worth noting that the spectra were all recorded on the same wavelengths and scale, and they do not contain missing values. Therefore we did not apply any preprocessing method. Before applying RT-RQDA we reduced the dimensionality by means of robust principal component analysis \citep{hubert2005robpca}, and selected $p=3$ components as they explain over $97.5\%$ of the variability. In the context of classification, there may be more appropriate techniques for dimensionality reduction such as approaches based on (local) linear discriminant analysis \citep{Martinez2001, Sugiyama2007}. However, it is recommended to use a robust method in order to avoid the loss of information in the reduced data due to the presence of outliers. Because of its simplicity and robustness and for the purpose of illustration, we decided on robust PCA.
Figure~\ref{fig:fruit_pca} shows the scores resulting from this dimensionality reduction step. We clearly see that class Ha consists of two subgroups, possibly due to the different illumination setups of the spectroscopy experiment. In addition we see that classes E and H may also contain subgroups.

\begin{figure}[!ht]
\centering
\vspace{0.2cm}
\includegraphics[width = 0.99\textwidth]{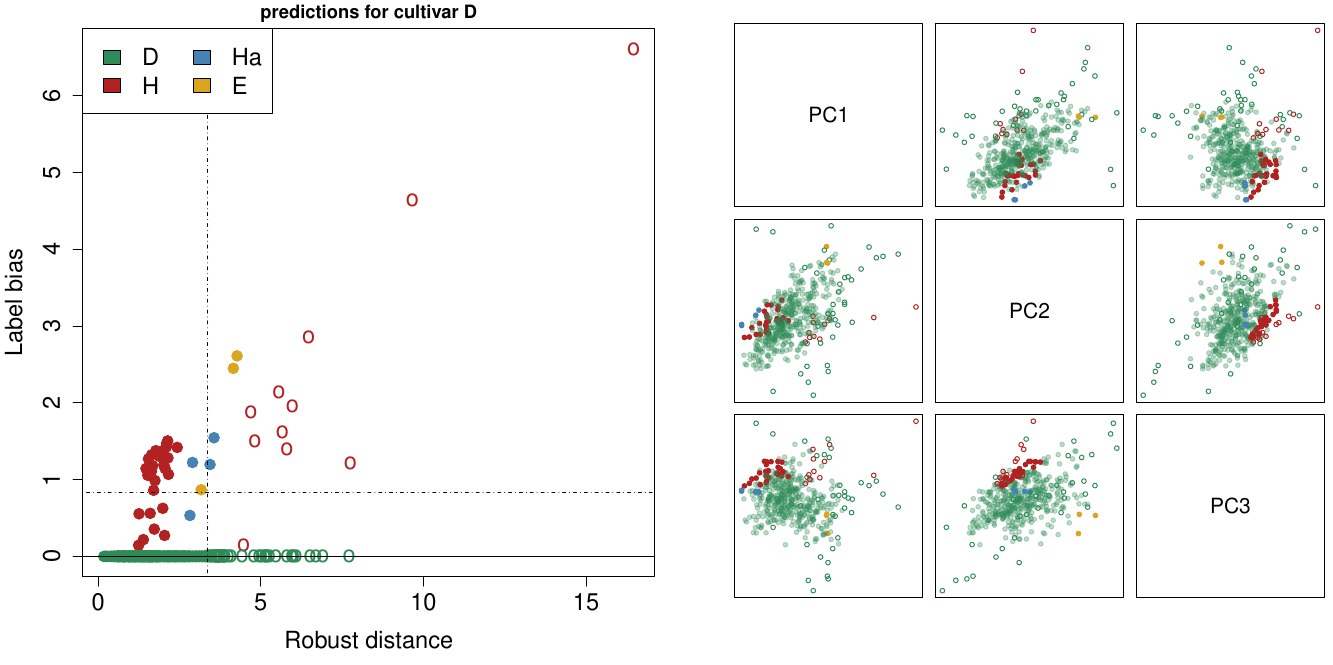}	
\caption{LB-plot of cultivar D (left) with a pairs plot of the PC scores of this class using the same colors and plotting symbols (right).}
\label{fig:fruit_LB1}
\end{figure}

We apply RT-RQDA to the data and construct the LB-plots of all classes.
Figure~\ref{fig:fruit_LB1} shows the LB-plot of cultivar D as well as a clarifying pairs plot containing only the points of cultivar D. All points in this LB-plot belong to class D, but some have LB $> 0$ meaning they are assigned to other classes indicated by their color. The red points in the LB-plot suggest that there is some overlap with class H. This overlap was to be expected from the original plot of the scores in Figure \ref{fig:fruit_pca}. In addition to the overlap with cultivar H, class D has quite a few outliers as can be seen from the green and red empty circles in the LB-plot. This is confirmed by the position of these points in the right panel of Figure~\ref{fig:fruit_LB1}, where the same colors and plotting symbols were used as in the left panel.

\begin{figure}[!ht]
\centering
\vspace{0.2cm}
\includegraphics[width = 0.99\textwidth]{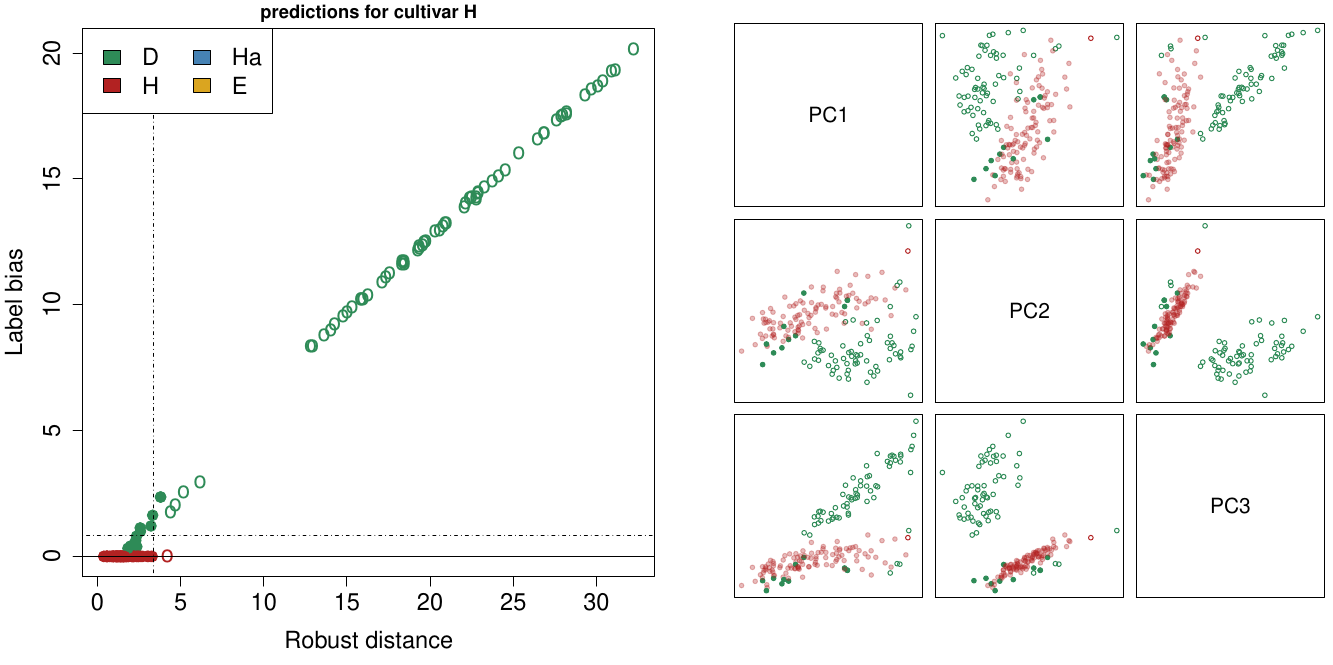}		
\caption{LB-plot of cultivar H (left) with a pairs plot of the PC scores of this class using the same colors and plotting symbols (right).}
\label{fig:fruit_LB2}
\end{figure}

The LB-plot of cultivar H is shown in Figure~\ref{fig:fruit_LB2}. We see right away that this cultivar consists of two subgroups. One subgroup is found in the bottom left corner of the LB-plot. This subgroup contains the bulk of the data, and has a little bit of overlap with cultivar D as indicated by the handful of green points with LB~$>0$. The other subgroup is found in the top right portion of the LB-plot. These green empty circles have a high label bias and a high robust distance. The latter indicates that they lie quite far from the bulk of the data, which is confirmed in the plot of (PC2,PC3) in the right panel of Figure~\ref{fig:fruit_LB2}. The fact that these objects are plotted as empty circles in the LB-plot indicates that they are also far from the other classes. It turns out that this subgroup corresponds exactly to the first 60 spectra measured for this cultivar. This strongly suggests that the experimental setup was changed after 60 measurements.

\begin{figure}[!ht]
\centering
\vspace{0.2cm}
\includegraphics[width = 0.99\textwidth]{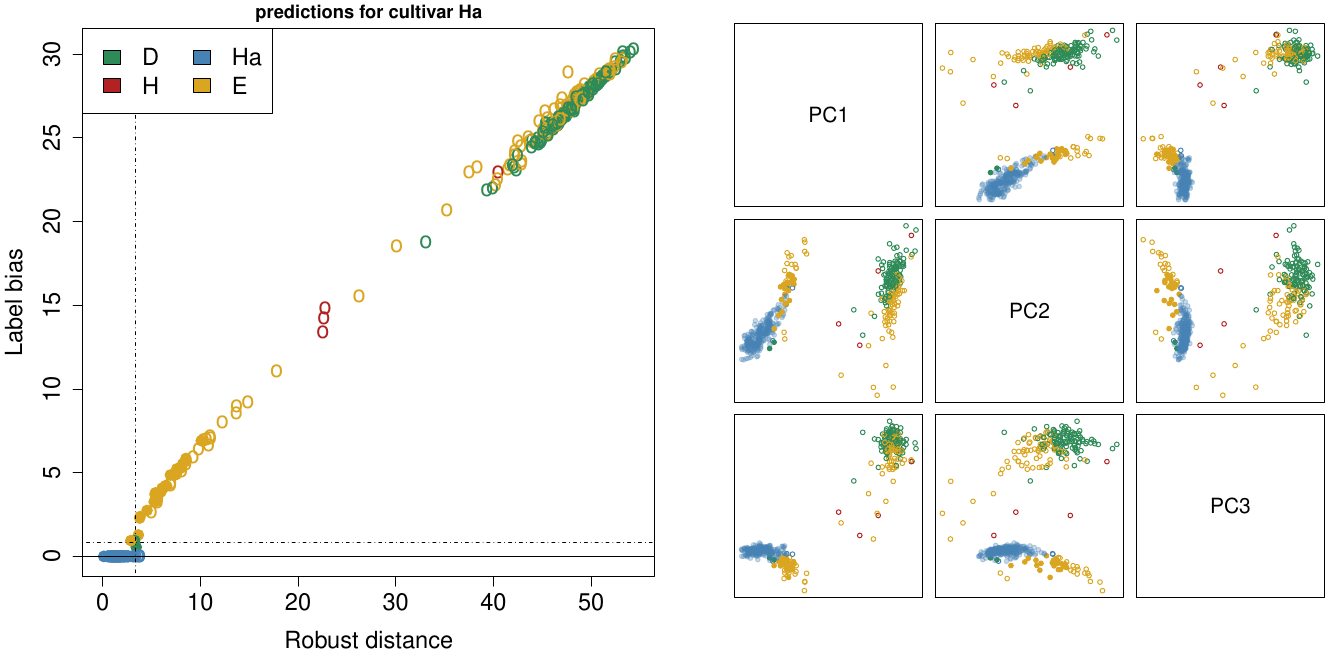}		
\caption{LB-plot of cultivar Ha (left) with a pairs plot of the PC scores of this class using the same colors and plotting symbols (right).}
\label{fig:fruit_LB3}
\end{figure}

Figure~\ref{fig:fruit_LB3} shows the LB-plot of cultivar Ha. It is known that during the spectroscopy experiments on this cultivar the illumination setup was changed twice. The initial illumination setup produced the first 180 points, which can be seen in the top right of the LB-plot. It corresponds to all points with a robust distance over 20. Therefore the initial illumination has produced a different data distribution from the others. This is confirmed by the clear separation in the pairwise plots in the right panel of Figure~\ref{fig:fruit_LB3}. The remaining 320 spectra of cultivar Ha are found in the bottom left of the LB-plot. The blue points are correctly assigned to class Ha, and the orange ones to class E. It is not immediately clear whether these orange points form a subgroup which should be considered separately, but in the pairs plots we see more of a gradual transition than a clean separation. 

\begin{figure}[!ht]
\centering
\vspace{0.2cm}
\includegraphics[width = 0.99\textwidth]{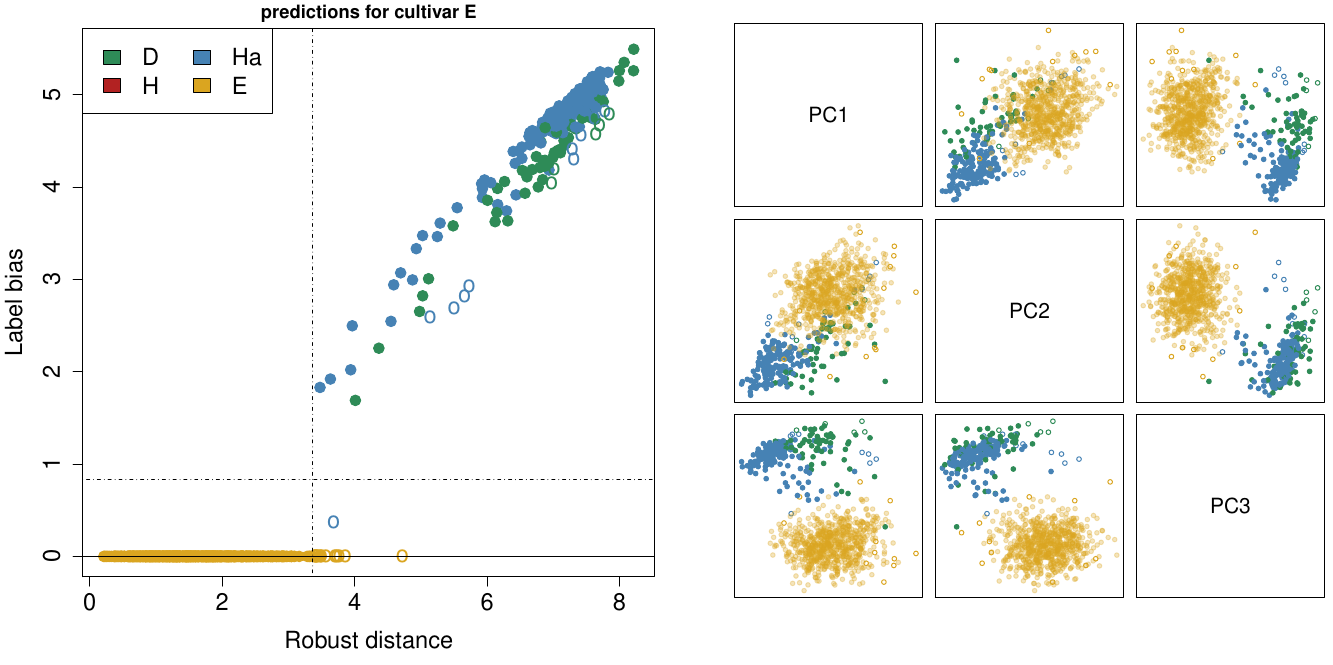}		
\caption{LB-plot of cultivar E (left) with a pairs plot of the PC scores of this class using the same colors and plotting symbols (right).}
\label{fig:fruit_LB4}
\end{figure}

The LB-plot of cultivar E is shown in Figure~\ref{fig:fruit_LB4}. Also here we see two distinct classes. The majority of the points has LB~$=0$ and a relatively small robust distance, suggesting a homogeneous subgroup. This is confirmed by the orange points in the right hand panel of Figure~\ref{fig:fruit_LB4}. The other subgroup is found in the top right portion of the LB-plot, and clearly separated from the bulk of the data in the pairwise plot of (PC2,PC3). It turns out that the spectra of this cultivar were collected on two different dates. The first 200 were measured roughly one month before the remaining 788. Furthermore, the outlying subgroup visible on the plot corresponds almost exactly with the first 200, suggesting that the subgroups cannot be considered as coming from the same population. This may be the result of different experimental setups, which is not unlikely given that the data of the two groups were collected one month apart.

In order to evaluate the classification performance of RT-RQDA on the fruit data, we split the data at random into a training set and a validation set containing 60\% and 40\% of the data (i.e.\ 1294 and 864 observations). Note that the validation set also contains outliers, which we do not want to take into account when evaluating classifiers. The points in the validation set that are outlying with respect to their class are therefore discarded in the calculation of misclassification rates. This procedure yields a validation set of 632 observations, with classes of sizes 179, 44, 106 and 303 for cultivars D, H, Ha and E. 

The left part of Table \ref{tbl:confmat_clas} shows the resulting confusion matrix of CQDA. Its misclassification rate is about 22\%. In the right part of Table~\ref{tbl:confmat_clas} we see that RT-RQDA performs better, with a misclassification rate of 2\%. The largest difference between CQDA and RT-RQDA occurs in the classification of cultivar Ha. Here CQDA assigns most non-outlying spectra to cultivar E, indicating that it has failed to characterize cultivar Ha due to its large number of outliers. The RT-RQDA method does classify cultivar Ha accurately, and only shows a slight confusion between cultivars D and H which is not surprising as these classes overlap.

\begin{table}[!ht]			
\centering
\caption{Confusion matrix of the validation set based on classical (left) and robust (right) discriminant analysis.}
\label{tbl:confmat_clas}
\vspace{0.3cm}
\resizebox{0.7\columnwidth}{!}{ 
\begin{tabular}{@{}lllcccccccccc@{}} 
\toprule[1.5pt]	
	&&& \multicolumn{4}{c}{CQDA} && \multicolumn{4}{c}{RT-RQDA} \\
	&& & {D}	& {H} & {Ha} & {E} && {D} &  {H} & {Ha} & {E} \\ 
\cmidrule{4-7} \cmidrule{9-12} 	
\noalign{\vskip 1mm} 
	&&D \phantom{fff}&  0.933  & 0.034   & 0.011   & 0.022   && 0.950    &0.050    &     0.000   &      0.000 \\
	&&H &  0.500  & 0.477   & 0.023   & 0.000   && 0.136    &0.864    &     0.000   &      0.000 \\
	&&Ha & 0.000  & 0.057   & 0.000   & 0.943   &&      0.000    &     0.000    &1.000   &      0.000 \\
	&&E &  0.000  & 0.000   & 0.000   & 1.000   &&      0.000    &     0.000    &     0.000   & 1.000 \\
\bottomrule
\end{tabular}
}
\end{table}

\section{Conclusion}\label{sec:conclusions}

Classical quadratic discriminant analysis (CQDA) is known to be sensitive to label noise and measurement noise (outliers). To address this issue we proposed a procedure for real-time robust quadratic discriminant analysis, called RT-RQDA. It incorporates the recent RT-DetMCD algorithm as well as an anomaly detection step, which flags observations that stand out relative to all classes. Robust discriminant scores are obtained from the RT-DetMCD estimates of location and scatter as well as robust membership probabilities. An extensive simulation study showed that the speed of RT-RQDA allows it to handle huge data sets and that it remains effective even if the data are contaminated by label and measurement noise simultaneously.

We also introduced a graphical display, the LB-plot, which gives insight into the presence of label and measurement noise in each class of the training data. The LB-plot makes atypical observations stand out, thereby identifying outliers, potential mislabeling, overlap, and observations which may be hard to classify. 

Finally, the proposed method was illustrated on two real datasets. In both applications, the RT-RQDA procedure with the LB-plot correctly identified several atypical observations.  

%\section*{Software availability}
%MATLAB implementations of our algorithms are available from the  website \url{http://wis.kuleuven.be/statdatascience/robust/software}\,.

\section*{Acknowledgements}
We thank Johan Speybrouck and Tim Wynants for their support throughout the project. We also acknowledge the financial support of VLAIO, grant HBC.2016.0208 (Development and implementation of real-time, robust statistical methods with novel applications in food sorting) for making this industrial research possible. The work of Mia Hubert, Jakob Raymaekers and Peter Rousseeuw was supported by the grant C16/15/068 (Statistical methodology for immense lively experiments) of the Research Council of KU Leuven.

%%%%%%%%%%%%%%%% Supplementary Material %%%%%%%%%%%%%%%%

\clearpage
\pagenumbering{arabic}
% restarts page numbering from 1
%\appendix
% \numberwithin{equation}{section} 
% restarts equation numbering from 1
\section*{Additional Material: Description of the\\ 
   RT-DetMCD algorithm}\label{app:RTDetMCD}
%\renewcommand{\theequation}
   %{\thesection.\arabic{equation}}
% labels equations as (A.1),...

We give a detailed description of the RT-DetMCD algorithm of \cite{de2020real} which computes the MCD estimator of \cite{rousseeuw1984least}. Given $n$ observations in $p$ dimensions, the objective is to find the subset $H$ of $h$ observations whose sample covariance matrix has the lowest possible determinant.

The raw MCD-estimate of location $\hbmu_H$ is the average of these $h$ points:
\begin{equation}
\hbmu_{H} = \frac{1}{h} \sum_{i \in H} \bx_i
\label{eq:rawcentering}
\end{equation}
whereas the raw scatter estimate is a multiple of their covariance matrix:
\begin{equation}
\hbSigma_{H} =  \frac{c_{\alpha}}{h-1} \sum_{i \in H } (\bx_i - \hbmu_h ) (\bx_i - \hbmu_h)^\top ,
\label{eq:rawcovar}
\end{equation}
where $c_{\alpha}$ is a consistency factor that depends on $h$ and $n$.\\
The best known algorithms for computing the MCD estimator are FastMCD \citep{rousseeuw1999fast} and DetMCD \citep{hubert2012deterministic}, both of which make use of so-called concentration steps (C-steps) to improve the subset $H$. Given a candidate subset of $h$ observations $H_0$, the C-step computes a more concentrated approximation by calculating the Mahalanobis distance of every observation $\bx_i$ based on the location and scatter of the current subset~$H_0$ as		
\begin{equation}
 \RD_i = \sqrt{(\bx_i - \hbmu_{H})^\top \hbSigma_{H}^{-1} (\bx_i - \hbmu_{H})}\,.
\label{eq:csteps}
\end{equation}		
These distances are then ordered, and the $h$ observations with the lowest distances form the new $h$-subset $H_1$. The new subset was proved to have an equal or lower determinant \citep{rousseeuw1999fast}, and this C-step procedure can thus be iterated until convergence. While guaranteed to converge, the procedure may only arrive in a local minimum. Therefore, the C-step procedure is run from different starting values, and after convergence the solution with the lowest determinant is retained. FastMCD starts from randomly sampled starting values, whereas DetMCD uses six deterministic starting values using several fast robust estimators in the first step. While FastMCD and DetMCD are popular algorithms that work well on medium sized datasets, their computational cost makes them impractical for real-time applications or very large industrial datasets. 

Recently, a parallel algorithm called RT-DetMCD was introduced \citep{de2020real}, which is suitable for very large datasets. In addition to the parallelization strategy, RT-DetMCD incorporates three key improvements: new initial estimators, fast C-step updates, and one-pass aggregation. The resulting algorithm is specifically designed for large datasets. It preserves the robustness and conceptual simplicity of the original MCD. As the algorithm is quite lengthy, we focus on the main differences compared with DetMCD. Further details are found in \citep{de2020real}. A research-level MATLAB implementation of RT-DetMCD is publicly available from the web page 
\url{http://wis.kuleuven.be/statdatascience/robust/software}.

The RT-DetMCD algorithm starts by splitting up the standardized observations $\bz_i$ into $q$ blocks $\bZ^{(\ell)}$ with $\ell=1,\dots, q$. For each block the algorithm computes the estimates 
$\hbmu_{raw}^{(\ell)}$ and $\hbSigma_{raw}^{(\ell)}$ by applying C-steps starting from two new initial estimators instead of DetMCD's six initial estimators.
These $q$ solutions then need to be aggregated in a robust way. They have many dimensions since the symmetric matrices $\hbSigma^{(\ell)}_{raw}$ contain $p(p-1)/2$ distinct entries, and the $\hbmu^{(\ell)}_{raw}$ have $p$ additional entries. Due to this high dimension, computing a typical robust estimate of the $q$ fits is problematic. Instead the entry-wise median of the $q$ fits is computed, yielding the
entry-wise median of the $\hbmu^{(\ell)}_{raw}$ denoted as
\begin{equation}
	\hbmu_{med} = 
	(\med_\ell((\hbmu^{(\ell)}_{raw})_1),\ldots,
	\med_\ell((\hbmu^{(\ell)}_{raw})_p)
\end{equation}
and the entry-wise median of all scatter matrices, given by 
\begin{equation}
(\hbSigma_{med})_{jk} =
\med_\ell((\hbSigma^{(\ell)}_{raw})_{jk})
\end{equation}
for $j,k=1,\ldots,p$.
However, as $\hbSigma_{med}$ is not necessarily positive definite one cannot use it as a final aggregated outcome. Instead, each thread computes the Kullback-Leibler deviation 
$\mbox{KL}[(\hbSigma_{med},\hbmu_{med}), 
(\hbSigma^{(l)}_{raw},\hbmu^{(l)}_{raw})]$ given by 
\begin{multline}
	\mbox{KL}[(\bA,\ba),(\bB,\bb)] \coloneqq 
	\text{trace}(\bA \bB^{-1} - \bI)
	- \ln|\bA \bB^{-1}| \\
	+ (\ba-\bb)^\top \bB^{-1}(\ba-\bb)\;.
	\label{eq:KL}
\end{multline}	
Next, the $h$-subsets of the blocks with the $[q/2]$ lowest KL deviations are aggregated by single-pass pooling.
% \citep{bennett2009numerically}. 
This yields the raw RT-DetMCD solution $(\hbSigma_{raw}, \hbmu_{raw})$. Each thread then computes squared distances $\RD^2(\bz_i,\hbmu_{raw},\hbSigma_{raw})$ and flags the $\bz_i$ for which they exceed a quantile of the chi-squared distribution with $p$ degrees of freedom. This yields a binary vector $\bdelta$ of length $n$ given by
\begin{equation*}
	\delta_i = \begin{cases}
		1 & \text{ if } \RD^2(\bz_i, \hbmu_{raw},
		     \hbSigma_{raw}) \leqslant \chi^2_{p, 0.975}\\
		0 & \text{ otherwise}.
\end{cases}
\end{equation*}
Then reweighted estimates are computed as the classical mean and covariance of the observations with $\delta_i = 0$, yielding the final RT-DetMCD estimates $(\hbSigma_{rew}, \hbmu_{rew})$. These estimates are then used one last time to flag the outliers in the dataset using the same rule as above. By undoing the initial standardization, all location and scale estimates are transformed back to the original coordinate system.

\end{document}